\documentclass{note}

\usepackage{fancyhdr}
\usepackage{setspace}
\usepackage{url}
\usepackage{graphicx}
\usepackage[colorlinks,linkcolor=blue,citecolor=magenta]{hyperref}
\usepackage{amsmath,amssymb,amsfonts,latexsym} 
\usepackage{authblk}
\usepackage[font=small,labelfont=bf]{caption}
\usepackage{boldline, multirow}
\usepackage{makecell}

\newtheorem{precor}{{\bf Corollary}}

\newtheorem{precon}{{\bf Conjecture}}

\newtheorem{predefin}{{\bf Definition}}

\newtheorem{preexm}{{\bf Example}}

\newtheorem{preappl}{{\bf Application}}

\newtheorem{prelem}{{\bf Lemma}}

\newtheorem{preproof}{{\bf Proof.\ }}

\newtheorem{presproof}{{\bf Sketch of Proof.\ }}

\newtheorem{prethm}{{\bf Theorem}}

\newtheorem{prealphthm}{{\bf Theorem}}

\newtheorem{prepro}{{\bf Proposition}}

\newtheorem{preprb}{{\bf Problem}}

\def\hom[#1,#2]{\text {${\rm Hom}({#1},{#2})$}}
\def\onvhom[#1,#2]{\text {${\rm Hom^{v}}(#1,#2)$}}
\def\onehom[#1,#2]{\text {${\rm Hom^{e}}(#1,#2)$}}

\def\authora{ \includegraphics[width=21mm]{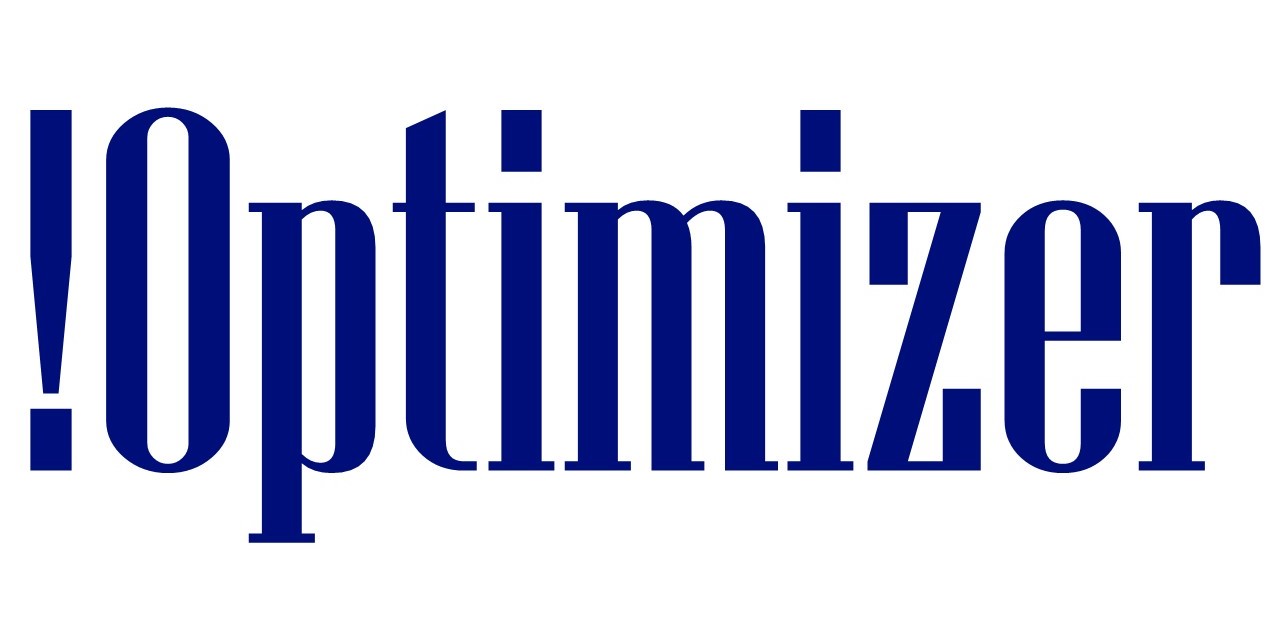}}
\def\authorb{}
\def\addressa{{\footnotesize Sharif Optimization and Applications Laboratory,\\
	          Department of Mathematical Sciences,
              Sharif University of Technology,\\
              Tehran, Iran.\\
              \url{optimizer.math.sharif.edu}}}
\def\addressb{}

\title{Multi-Feasibility Variable Selection}

\author[$\dagger$]{Ali Fathi}
\author[$\dagger$]{Mohammad Rashid}
\author[$\dagger$]{Shayan Ranjbarzadeh}
\author[$\dagger$]{Mojtaba Tefagh 
\thanks{Correspondence: \href{mailto:mtefagh@sharif.edu}{mtefagh@sharif.edu}}}

\affil[$\dagger$]{Sharif Optimization and Applications Laboratory, Department of Mathematical Sciences, Sharif University of Technology}

\date{}

\begin{document}

\renewcommand{\headrulewidth}{.4pt}
\lhead{\sc !Optimizer 2021 Competition Report}\chead{} \rhead{\thepage}\lfoot{}\cfoot{}\rfoot{}
\pagestyle{fancy}

\maketitle
 
 \begin{abstract}
This paper is the report of the problem 
proposed for the \href{http://optimizer.math.sharif.edu/Optimizer2021.pdf}{\sc !Optimizer 2021} competition, and the solutions of the gold medalist team, \emph{i.e.}, the \href{https://github.com/Optimizer-Competition-Pandas}{Panda} team. 
The competition was held in two stages, the research and development stage and a two-week contest
stage, consisting of five rounds, and seven teams succeeded in finishing both stages to the end. In this joint report of the winner team Panda and the problem design committee coordinated by Mojtaba Tefagh,
 we first explain each of the five rounds and then provide the solutions proposed by our team (Panda) to fulfill the required tasks in the fastest and most accurate way. 
Afterward, some preprocessing and data manipulating ideas used to enhance the algorithms would be presented. All codes are written in the Julia language, 
which showed a better performance than Python on optimization problems in our comparisons during the R\&D stage, and are publicly available in this 
Github \href{https://github.com/Optimizer-Competition-Pandas}{repository}.
 \end{abstract}

\section{Competition Motivations and Background}
In inverse optimization models, one wants to learn the parameters of a family of optimization problems in a way that some desirable points would become optimal for their associated instances \cite{YANG199969}. In this contest, we consider the following class of parameterized feasibility \emph{linear programs} (LP) extensively studied in the field of \emph{constraint-based reconstruction and analysis} (COBRA)
\begin{equation}\label{FBA}
  \begin{array}{ll}
  \text{find} & v \\
  \text{subject to} & S^{I}v = 0, \\
  & l^{I} \preceq v \preceq u^{I}, \\
  & I=\{i_1,i_2,\ldots,i_k\}\subseteq \{1,2,\ldots,n\}.
  \end{array}
\end{equation}
where $S=[S_1,S_2,\ldots,S_n]$ is an $m \times n$ stoichiometric matrix whose columns $S_i$ characterize the stoichiometry of the biochemical reactions happening inside a cell,
\begin{align*}
    l = \begin{bmatrix}
           l_{1} \\
           l_{2} \\
           \vdots \\
           l_{n}
         \end{bmatrix}
   \quad\text{and}\quad
    u = \begin{bmatrix}
           u_{1} \\
           u_{2} \\
           \vdots \\
           u_{n}
         \end{bmatrix}
\end{align*}
are lower and upper bound vectors of dimension $n$ as determined by the growth media capacity and thermodynamic constraints, while the subvectors 
\begin{align*}
    l^{I} = \begin{bmatrix}
           l_{i_1} \\
           l_{i_2} \\
           \vdots \\
           l_{i_k}
         \end{bmatrix}
   \quad\text{and}\quad
    u^{I} = \begin{bmatrix}
           u_{i_1} \\
           u_{i_2} \\
           \vdots \\
           u_{i_k}
         \end{bmatrix}
\end{align*}
as well as the submatrix $S^{I}=[S_{i_1},S_{i_2},\ldots,S_{i_k}]$ contain the corresponding entries and columns, respectively.

A metabolic network comprises the collection of all chemical pathways of the metabolism of an organism, \emph{e.g.}, a bacterium. In systems biology, it is observed that if we reconstruct a metabolic network by determining the set of indices $I$ representing its reactions, whether the associated organism is viable or not under circumstances specified by $l$ and $u$  can, to some extent, be predicted by whether LP \eqref{FBA} is feasible or not \cite{orth2010flux}. Moreover, numerous experiments have demonstrated the consistency of cell viability predictions derived by the feasibility of  LP \eqref{FBA} with wet lab measurements \cite{lewis2012constraining}. 

Turning this argument around, one may try to reconstruct a genome-scale metabolic network, given a list of viable and nonviable scenarios, by exploiting the fact that all the solutions to different instances of  LP \eqref{FBA} have the same sparsity pattern because they belong to different strains of the same species \cite{10.1371/journal.pcbi.0020072,10.1371/journal.pcbi.1000308}. Therefore, one may select the candidate $I$ by utilizing methods based on joint group sparsity. Other possible applications of this developed framework include, but are not limited to, signal processing \cite{HEBO14}, astrophysics \cite{Starck_2016}, photoplethysmography \cite{FOBA18,FOBA18-2}, and inverse scattering problem \cite{SUKY18,SUDX20}.


\section{Problem Formulation and Methods}
The competition task was broken into five gradually complicating rounds, with the main sparse reconstruction problem at the last round. In this section, we present the problem 
formulations of each round and provide our methods and algorithms for them, from the first round to the fifth.
\subsection*{Round $1$: Steady-state flux distributions}\label{r:1}
	The $s_{ij}$ entry of $S$ represents the molar rate of either consumption (if $s_{ij} \leq 0$) or production (if $s_{ij} \geq 0$) of the metabolite $i$ in the reaction $j$ per unit of dry 
	cell weight. If all the metabolites are in mass balance at specific concentrations, \emph{i.e.}, $Sv = 0$, we say that the metabolic network is in the steady-state condition. 

	In this round, we were supposed to find the vector $v$ of the rates of reactions subject to the constraints
	\[
		Sv = 0,
		\qquad
		l^1 \preceq v \preceq u^1,
	\]
	as predicted by \emph{flux balance analysis} (FBA) \cite{10.1093/bib/bbl007}, \emph{i.e.},
	\[
  		\begin{array}{ll}
  			\text{find} & v \\
  			\text{subject to} & Sv = 0, \\
  			& l^{1} \preceq v \preceq u^{1}.
 			 \end{array}
	\]

	To fulfill this task, we simply used a usual LP optimization code with objective function set to $0$, \emph{i.e.}, $\text{maximize } 0^Tv$. 
	Also, the code and the answer produced by the next round could be used here as a feasible vector. Our Julia code for this round, placed in a jupyter notebook file, could 
	be found \href{https://github.com/Optimizer-Competition-Pandas/Round_1/blob/main/Optimizer_R1.ipynb}{here}.

\subsection*{Round $2$: Convex relaxation of cardinality optimization problems}\label{r:2}
	In this round, we were required to find the most sparse flux vector satisfying the constraints
	\begin{equation}
		Sv = 0,
		\qquad
		l^1 \preceq v \preceq u^1,
	\end{equation}
	by minimizing $\| v \|_{0}$, \emph{i.e.},
	\begin{equation}
 	 	\begin{array}{ll}
  			\text{minimize} & \| v \|_{0} \\
  			\text{subject to} & Sv = 0, \\
  			& l^{1} \preceq v \preceq u^{1},
 		\end{array}
	\end{equation}
	and the biological intuition behind the theory is to minimize the total enzyme load imposed on the organism \cite{maranas2016optimization}.
	\\
	\indent To do so, some different algorithms were tested in this round (such as the dual-density method \cite{Xu2021} and five other algorithms), 
	but the most successful one was a method called \emph{weighted $l_1$-norm minimization} 
	\cite{Zhao_2018}, which we explain here in detail. The weighted algorithm optimizes 
	the following problem:
	\begin{equation}
		\label{eq:w}
  		\begin{array}{ll}
  			\text{minimize} & \sum_{i=1}^{n} w_{i} | v_{i} | \\
  			\text{subject to} & Sv = 0, \\
  			& l^{1} \preceq v \preceq u^{1}.
 			 \end{array}
	\end{equation}

	This objective function, the weighted sum of the absolute values of the elements of $v$, is a generalization of approximating $l_0$-norm by $l_1$-norm.
	Substituting $w$ with $\vec{1}_n$, the weighted problem would be simply the $l_1$-norm minimization. 
	Also, defining a set of zero indices $I_z$ and putting $w_i \approx \infty$ for each $i \in I_{z}$, and $w_i=0$ otherwise, makes this problem to find some sparse solution 
	in which $v_{I_z} = 0$, which implies that this formulation includes all sparse optimization problems of interest by having the appropriate weights. \\
	To find the most proper weights, we use an iterative algorithm. The weights are iteratively updated according to the optimal solution $v$ of the previous step, and the current problem with these 
	weights would be solved to get a new optimal vector $v$ and so it continues. There are many possible update rules for $w$; for instance, a choice for converging to a sparse result could be  
	like the following rule: 
	\begin{align}
		& w^{(0)} = \vec{1}_n \nonumber \\
		& w_i^{(t+1)} = \frac{1}{|v_i^{(t)}| + \epsilon} \label{w:1}
	\end{align}
	By this rule, after solving the problem on step $t$ to get $v^{(t)}$, weights for the next iteration are defined inversely related to the magnitude of elements of $v^{(t)}$. If $v^{(t)}_i$ is small for
	some index $i$, its corresponding weight in the next iteration, $w_i^{(t+1)}$, is set to a large amount, to try to force it to zero (if possible). The intuition behind it is that if some 
	elements of $v$ are near to $0$, reducing them by some $\epsilon$ and compensating this reduction by the elements of $v$ with large magnitudes, has probably an insignificant effect on 
	$l_1$-norm (and the approximate objective function) but a significant effect on $l_0$-norm, which is increased only when some elements of $v$ are set to 
	\emph{absolute zero}. Note that $\epsilon$ prevents numerical issues (such as division by zero) and was set to $10^{-5}$ in our best practice result (we will discuss
	later that $\epsilon$ plays some other roles in the theoretical analysis).  \\
	
	A variation of rule \eqref{w:1} was used in our final solution, named NW4 \cite{Xu2021}, which is as follows
	\begin{align}
		& w^{(0)} = \vec{1}_n, \nonumber \\
		& w_i^{(t+1)} = \frac{1 + (|v_i^{(t)}| + \epsilon)^{p}}{(|v_i^{(t)}| + \epsilon)^{p+1}} \label{w:4},
	\end{align}
	in which $p$ is some modifiable parameter and was epmirically set to $0.8$ to get the best result on the competition's data. By setting $p$ to $0$, this rule would be 
	the same with rule \eqref{w:1} (up to a constant $2$ which makes no change in the objective function).
	\\
	However, NW4 was not the final update rule we used. As mentioned earlier, this converging-to-sparse rule is just a heuristic and could fall into some local optima. 
	To prevent this issue, we added some randomness to the algorithm in the following way:
	\begin{align*}
		& w^{(0)} = \vec{1}_n \\
		& w_i^{(t+1)} = \frac{1 + (|v_i^{(t)}| + \epsilon)^{p}}{(|v_i^{(t)}| + \epsilon)^{p+1}} \times r_i^3 \\
		& r_i \sim Unif[0,1]
	\end{align*}
	The distribution of randomness and the way it has appeared in $w$ are set empirically to make the best results. \\
	
	The mentioned updating rules \eqref{w:1} and \eqref{w:4} are not merely heuristics. The theory behind the weighted algorithm is using some convex (or concave) function to
	approximate $l_0$-norm, called merit function, $\Phi_{\epsilon}(v)$ such that:
	\begin{equation}
		\label{eq:merit}
		\lim_{\epsilon \rightarrow 0} \Phi_{\epsilon}(v) = \left \lVert v \right \rVert_0
	\end{equation} 
		In general, $\Phi_{\epsilon}$ is a convex function to make the final problem a convex one. However, $\Phi_{\epsilon}$ is sometimes chosen to be a non-convex function, 
	but is then approximated by a linear function afterward (which transforms the original problem to LP):
	\begin{equation}
		\label{eq:linapprox}
		\Phi_{\epsilon}(v) \approx \Phi_{\epsilon}(v^{(t)}) + {\nabla \Phi_{\epsilon}(v^{(t)})}^{T}.(v-v^{(t)})
	\end{equation}
	
	For instance, one famous choice for approximating $l_0$-norm is by the logarithmic function
	\begin{equation*}
		\Phi_{\epsilon}(v) = \sum_{i=1}^n \log{(|v_i| + \epsilon)},
	\end{equation*}
	which satisfies property \eqref{eq:merit}. Approximating this $\Phi_{\epsilon}$ with a linear function results in the following weights
	\begin{equation*}
		w^{(t+1)} = \nabla \Phi_{\epsilon}(v^{(t)}) =
		\Bigg( \frac{1}{|v_1^{(t)}|+\epsilon}\ , \cdots\ , \frac{1}{|v_n^{(t)}|+\epsilon} \Bigg)^{T},
	\end{equation*}
	which is exactly rule \eqref{w:1}. The merit function $\Phi_{\epsilon}$ for NW4 rule has a few variations which could be found at \cite{Xu2021}, and we don't include the details here.
	
\subsection*{Round $3$: Exact multi-feasibility variable selection}\label{r:3}
	The goal of this round, is to find the unknown matrix $V$  with jointly sparse columns which satisfies the following constraints
	\[
	SV = 0,
	\qquad
	L \preceq V \preceq U.
	\]
	Joint sparsity for an arbitrary set of sparse vectors means that all members of the set share a common sparse support set, \emph{i.e.},
	\begin{equation}\label{MMV}
  		\begin{array}{ll}
  			\text{minimize} & \| V \|_{2,0} \\
  			\text{subject to} & SV = 0, \\
  			& L \preceq V \preceq U,
  		\end{array}
	\end{equation}
	where the mixed norm is defined as follows
	\[
   		\| X \|_{p,q} = \| (\| x^{\prime}_1 \|_p, \| x^{\prime}_2 \|_p, \ldots, \| x^{\prime}_m \|_p) \|_q.
	\]

	To the end of this paper, we assume that $x^{\prime T}_1, x^{\prime T}_2, \ldots, x^{\prime T}_m$ are the rows of $X$ in this definition, but some authors use another convention of considering the columns 
	instead of rows. Apart from the difference in notation, the two definitions become clearly equivalent to one another if applied to the transpose matrix.

	To solve this problem, we approximated $l_{2,0}$-norm with $l_{1,1}$-norm
	\begin{equation}
		\label{eq:n11}
  		\begin{array}{ll}
  			\text{minimize} &  \left \lVert V \right \rVert_{1,1} = \sum_{j=1}^{c} \left \lVert v_{j} \right \rVert_1 \\
  			\text{subject to} & Sv_j=0  \quad \forall j, \\
  			& l_{j} \preceq v_{j} \preceq u_{j} \quad \forall j,
  		\end{array}
	\end{equation}
	in which $c$ is the number of columns in $V$ and $v_j$ denotes the $j$-th column of $V$. \\
	To justify this approximation, first, we see that in $\left \lVert V \right \rVert_{p,0}$, the $p$-norm of rows together with the $l_0$-norm, determines whether a row is all-zero or not.
	Both $p=1$ and $p=2$ (and other values for $p$) fulfill this job. In fact, we have exact equality, \emph{i.e.}, 
	$\left \lVert V \right \rVert_{1,0} = \left \lVert V \right \rVert_{2,0}$. 
	Afterward, $l_0$-norm in $\left \lVert V \right \rVert_{1,0}$ is replaced with $l_1$-norm to get $\left \lVert V \right \rVert_{1,1}$. To emphasize, it could be mentioned that the
	structure of rows and columns in $l_{2,0}$-norm is not maintained well, but as discussed, this approximation could be better understood when viewed as the combination of two steps, \emph{i.e.}, relaxing $l_2$-norm by $l_1$-norm and then $l_0$-norm by $l_1$-norm. Besides, we will replace $l_1$-norm
	approximation by a similar weighted sum as in \eqref{eq:w} to maintain the $l_0$-norm structure more precisely. Also, the successes of these approximations are demonstrated in
	practice.  \\
	Problem \eqref{eq:n11} is LP and can be consequently solved efficiently. However, there are some other advantages to this form. Namely, this problem could be separated into 
	$c$ independent problems
	\begin{equation}
  		\begin{array}{ll}
  			\text{minimize} & \left \lVert v_{j} \right \rVert_1 \\
  			\text{subject to} & Sv_j=0, \\
  			&  l_{j} \preceq v_{j} \preceq u_{j} ,
  		\end{array}
	\end{equation}
	for $1 \leq j \leq c$. Separation helps in the case that if the solver is super-linear (\emph{e.g.}, $O(n^{1+\delta})$ for an arbitrary $\delta$), 
	having $c$ distinct problems of size $n$ would be solved faster than a problem of size $c \times n$.
	
	As mentioned before, we need to make some changes to \eqref{eq:n11} to keep the $l_0$-norm structure in the $l_{2,0}$-norm. It could be done by applying the weighted algorithm,
	exactly like \eqref{eq:w}. The modified problem would be as the following
	\begin{equation}
  		\begin{array}{ll}
  			\text{minimize} & \sum_{i=1}^{n} w_{i} \left \lVert v^{\prime}_{i} \right \rVert_1 \\
				  	      &= \sum_{j=1}^{c} (\Sigma_{i=1}^{n} w_{i} |{(v_j)}_i|) \\
  			\text{subject to} & Sv_{j}=0 \quad \forall j, \\
  			&  l_{j} \preceq v_{j} \preceq u_{j} \quad \forall j,
  		\end{array}
	\end{equation}
	in which $v^{\prime}_i$ denotes the $i$-th row of $V$. As the problem has remained linear, it could be separated again as in the following: 
	\begin{equation}
  		\begin{array}{ll}
  			\text{minimize} & \sum_{i=1}^{n} w_{i} |{(v_j)}_i| \\
  			\text{subject to} & Sv_j=0, \\
  			&  l_{j} \preceq v_{j} \preceq u_{j} ,
  		\end{array}
	\end{equation}
	The weights are updated just the same as in the algorithm for round \hyperref[r:2]{2}, but by substituting $\left \lVert v_i^{\prime(t)} \right \rVert_2$ or $\left \lVert v_i^{\prime(t)} \right \rVert_1$
	instead of $|v_i^{(t)}|$. For example, for rule \eqref{w:1}:
	\begin{align}
		& \vec{w^{(0)}} = \vec{1}_n \nonumber \\
		& w_i^{(t+1)} = \frac{1}{\left \lVert v_i^{\prime(t)} \right \rVert_2 + \epsilon} 
	\end{align}
	The code for this round is available \href{https://github.com/Optimizer-Competition-Pandas/Round_3/blob/main/Optimizer_R3.ipynb}{here}.

\subsection*{Round $4$: Multi-feasibility variable selection in the presence of error}\label{r:4}
	In this round, it is requested to find the unknown matrix $V$  with jointly sparse columns when the matrix $SV$ is constrained to have jointly sparse rows and 
	$L \preceq V \preceq U$, \emph{i.e.},
	\begin{equation}
  		\begin{array}{ll}
  			\text{minimize} & \big(\| V \|_{2,0}, \| (SV)^T \|_{2,0}\big) \\
  			\text{subject to} & L \preceq V \preceq U.
 		\end{array}
	\end{equation}
	Here in fact, this multi-criterion objective is meant to be interpreted as follows \cite{boyd2004convex}:
	\begin{equation}
  		\begin{array}{ll}
  			\text{minimize} & \| V \|_{2,1} + \lambda \| (SV)^T \|_{2,1} \\
  			\text{subject to} & L \preceq V \preceq U.
 		\end{array}
	\end{equation}

	In other words, it indicates that freeing every $Sv_j=0$ equation would make a $\lambda$ penalty. Similar to the previous round, weighted algorithm and separation are utilized to solve this 
	multi-columns problem, but this time some $Sv_j=0$ equations are freed:
	\begin{equation}
  		\begin{array}{ll}
  			\text{minimize} & \sum_{i=1}^{n} w_{i} \left \lVert v^{\prime}_{i} \right \rVert_1 \\
  			\text{subject to} & Sv_{j}=0 \quad \forall j \in J, \\
  			&  L \preceq V \preceq U,
  		\end{array}
	\end{equation}
	First, we calculate $c$ variables $d_1, \ldots, d_c$ to determine a proper set $J$, which is meant to represent the constraints to be satisfied:
	\begin{align}
		d_{j} =& (\text{min  } \left \lVert v_{j} \right \rVert_1 \text{ s.t. } Sv_{j}=0 \text{ and }  l_{j} \preceq v_{j} \preceq u_{j}) \nonumber \\
			 &- (\text{min  } \left \lVert  v_{j} \right \rVert_1 \text{ s.t. }  l_{j} \preceq v_{j} \preceq u_{j} ) \label{heu_d}
	\end{align}
	Here $d_j$ is a heuristic of the advantage gained by freeing column $j$, measured by the fall in $l_1$-norm. Then, if this proxy of benefit suggests an improvement
	more than $\lambda$, we would free its corresponding constraint, \emph{i.e.},
	\begin{equation}
		\label{eq:J}
  		J = \{ j \, | \, d_j < \lambda \}.
	\end{equation}
	Every other detail is exactly similar to the previous round. The code for this round is available 
	\href{https://github.com/Optimizer-Competition-Pandas/Round_4/blob/main/Optimizer_R4.ipynb}{here}.
	
\subsection*{Round $5$: Multi-feasibility/infeasibility variable selection}\label{r:5}
	Making some small changes in the previous task, in this round we are going to solve \eqref{MMV} with the additional constraint that at most $K$ columns of $SV$ may have 
	nonzero entries, \emph{i.e.},
	\begin{equation}\label{Recon}
  		\begin{array}{ll}
 			\text{minimize} & \| V \|_{2,0} \\
  			\text{subject to} & \| (SV)^T \|_{2,0} \leq K, \\
  			& L \preceq V \preceq U.
  		\end{array}
	\end{equation}
	To address the importance and the biological intuition of this formulation, suppose that $\tilde{L}$ and $\tilde{U}$ have $t$ columns denoted by 
	$\tilde{l}_1,\tilde{l}_2,\ldots,\tilde{l}_t$ and $\tilde{u}_1,\tilde{u}_2,\ldots,\tilde{u}_t$, respectively. Consider the following feasibility problems for $1\leq k \leq t$
	\begin{equation}\label{infeasibility}
 		 \begin{array}{ll}
  			\text{find} & v \\
  			\text{subject to} & S^{I}v = 0, \\
  			& \tilde{l}_k^{I} \preceq v \preceq \tilde{u}_k^{I},
  		\end{array}
	\end{equation}
	where $I$ is defined as follows
	\begin{equation}
		I=\{i \mid \max_j \lvert V_{ij} \lvert > 0\}.
	\end{equation}

	According to the biological model, we know that these feasibility problems should be infeasible for the ground truth $I$, which represents the underlying metabolic network, and the lower and upper bounds 
	$\tilde{L}$ and $\tilde{U}$, which represent the different growth environments or other conditions. 
	\newline
	\indent Therefore, we will validate each solution $V$ by the percentage of the infeasible instances of \eqref{infeasibility} for the corresponding $I$. Note that, solving 
	\eqref{Recon} helps to get a better score since we know \emph{a priori} that the smaller the set of indices $I$, the higher the probability of infeasibility for each LP of the form 
	\eqref{infeasibility}.
	
	Our proposed method to solve \eqref{Recon} is similar to the one in round \hyperref[r:4]{4}. The problem that we solve is as follows:
	\begin{equation}
  		\begin{array}{ll}
 			\text{minimize} &\sum_{i=1}^{n} w_{i} \left \lVert v^{\prime}_{i} \right \rVert_1 \\
  			\text{subject to} & Sv_{j}=0 \quad \forall j \in J, \\
  			& L \preceq V \preceq U.
  		\end{array}
	\end{equation}
	 Again, the heuristic coefficients are defined similarly:
	\begin{align}
		d_{j} =& (\text{min  } \left \lVert v_{j} \right \rVert_1 \text{ s.t. } Sv_{j}=0 \text{ and }  l_{j} \preceq v_{j} \preceq u_{j})  \nonumber \\
			 &- (\text{min  } \left \lVert  v_{j} \right \rVert_1 \text{ s.t. }  l_{j} \preceq v_{j} \preceq u_{j} ) 
	\end{align}
	This time, we free the $K$ most advantageous columns, \emph{i.e.},
	\begin{equation}
  		J = \{ j \, | \, d_j < K\text{-th maximum coefficient in } d_1, \ldots, d_c \}
	\end{equation}
	The code for this round is available 
	\href{https://github.com/Optimizer-Competition-Pandas/Round_5/blob/main/Optimizer_R5.ipynb}{here}.

\section{Preprocessing and Data manipulations}
	To make our codes more efficient and faster, some data modifications have been used. We review the most important ones of them:
	\begin{description}
		\item[Ignoring non-zero elements:] In round \hyperref[r:2]{2}, if for some elements of $v$ like $v_i$ we have $0 < l_i$ ($\leq v_i$) or $0 > u_i$ ($\geq v_i$), then it is 
		intrinsically impossible to have $v_i=0$. Therefore, it is mishandling to try to make it zero (as the effort of increasing sparsity). Thus, $v_i$ could be taken out
		from the sparsity objective function to give the model more freedom over these elements. The same matter exists in rounds \hyperref[r:3]{3}, \hyperref[r:4]{4}, and \hyperref[r:5]{5},
		 where we have
		$L \preceq V \preceq U$ and if there is any element $L_{i,j} > 0$ or $U_{i,j}<0$, as it forces $V_{i.j}\neq 0$, it would be impossible for that row $i$ to play a role
		in joint-sparsity and therefore, this row could be taken out from the joint-sparsity objective function. In a more formal way:
		\begin{equation}
			\forall i [\exists j \left[ (L_{i,j}>0) \lor (U_{i,j}<0) \right] \Rightarrow w_i = 0]
		\end{equation}
		\item[Setting definite elements:] If for some elements we have $l_i=u_i$ ($L_{i,j}=U_{i,j}$ in the last rounds), $v_i$ ($V_{i,j}$) would be set to that definite value too, 
		and there is no need to contain that variable in our optimization problem. In the case of the data for this competition, this equaled value has always been $0$
		(\emph{i.e.}, $l_i = 0 = u_i$ or $L_{i,j} = 0 = U_{i,j}$), and setting $v_i$ ($V_{i,j}$) to zero, is like deleting those variables from the problem without any further effort. 
		These deletions have made the size of the problems in this competition significantly smaller and have caused the running time to drop considerably.
		\item[Sparsity lower-bound analysis:] As mentioned, those rows in which $L_{i,j} > 0$ or $U_{i,j}<0$ induce $V_{i,j}\neq 0$ and therefore define a lower bound for
		$\left \lVert V \right \rVert_{2,0}$. This lower bound could be modified. Let the algorithm result be $\bar{v}$, and $I_{nz}$ be the set of indices of all non-zero elements of $\bar{v}$.
		Roughly speaking, it is expected that for most of those $i \in I_{nz}$, forcing $v_i$ to zero would cause the problem to get infeasible. As $|I_{nz}|$ is comparatively small,
		we can test it by checking $|I_{nz}|$ feasibility problems. If knocking-out $i \in I_{nz}$ maked the problem infeasible, we can take $w_i=0$ and increase our lower bound.
		Doing so in the competition has ensured us that our final results are adequately close to optimal, as the distance between our $\left \lVert V \right \rVert_{2,0}$'s and the calculated
		lower bounds have been satisfactorily small.
	\end{description}
\section{Results}
	In this \href{https://github.com/mtefagh/Optimizer}{link} created by the organizer committee, you can find $7$ datasets for each round of the competition. These datasets consist of the
	metabolic models of the organisms Escherichia Coli, Salmonella, Cricetulus Griseus, Phaeodactylum Tricornutum, Mus Musculus, Homo Sapiens, and the BiGG Universal Model,
	increasing by size respectively. In the actual competition, $3$ of these datasets for each round were requested to be worked on. The performance of our codes on this comprehensive
	$7 \times 5$ datasets are summarized in the following table:
	\begin{center}
    	\begin{tabular}{ V{2.5} c V{2.5} c | c | c | c | c | c V{2.5}}
   		\Xhline{2\arrayrulewidth}
    		Round & Dataset & $m$ & $n$ & $c$ & s.c. & Running Time \\ \Xhline{2\arrayrulewidth}
    		\multirow{7}{*}{$1$} & E. Coli 					& $72$ 		& $95$ 		& $1$ 	& -  		& $0.119 \pm 0.270$ ms \\ \cline{2-7}
					       & Salmonella 				& $2436$ 		& $3357$ 		& $1$ 	& -  		& $7.563 \pm 3.607$ ms \\ \cline{2-7}
					       & P. Tricornutum 			& $2172$ 		& $4456$ 		& $1$ 	& -  		& $10.16 \pm 3.57$ ms \\ \cline{2-7}
					       & C. Griseus 				& $4456$ 		& $6663$ 		& $1$ 	& - 		& $25.44 \pm 9.85$ ms \\ \cline{2-7}
					       & Mus Musculus 			& $8404$ 		& $13094$ 	& $1$ 	& -  		& $72.02 \pm 33.97$ ms \\ \cline{2-7}
					       & Homo Sapiens 			& $8399$ 		& $13543$	& $1$ 	& -  		& $71.92 \pm 32.04$ ms \\ \cline{2-7}
		 			       & Universal Model 			& $15638$ 	& $28301$ 	& $1$ 	& -  		& $0.161 \pm 0.003$ s \\ \Xhline{2\arrayrulewidth}
    		\multirow{7}{*}{$2$} & E. Coli 					& $72$ 		& $95$ 		& $1$ 	& $8$  	& $3.221 \pm 0.505$ ms \\ \cline{2-7}
					       & Salmonella 				& $2436$ 		& $3357$ 		& $1$ 	& $5$  	& $47.15 \pm 4.25$ ms \\ \cline{2-7}
					       & P. Tricornutum			 	& $2172$ 		& $4456$ 		& $1$ 	& $93$  	& $0.125 \pm 0.007$ s \\ \cline{2-7}
					       & C. Griseus 				& $4456$ 		& $6663$ 		& $1$ 	& $95$ 	& $0.193 \pm 0.012$ s \\ \cline{2-7}
					       & Mus Musculus 			& $8404$ 		& $13094$ 	& $1$ 	& $101$  	& $0.428 \pm 0.027$ s \\ \cline{2-7}
					       & Homo Sapiens 			& $8399$ 		& $13543$ 	& $1$ 	& $106$  	& $0.482 \pm 0.040$ s \\ \cline{2-7}
		 			       & Universal Model 			& $15638$ 	& $28301$ 	& $1$ 	& $514$  	& $1.055 \pm 0.042$ s \\ \Xhline{2\arrayrulewidth}
    		\multirow{7}{*}{$3$} & E. Coli 					& $72$ 		& $95$ 		& $20$ 	& $9$  	& $0.249 \pm 0.005$ s \\ \cline{2-7}
					       & Salmonella 				& $2436$ 		& $3357$ 		& $20$ 	& $53$  	& $14.96 \pm 0.15$ s \\ \cline{2-7}
					       & P. Tricornutum			 	& $2172$ 		& $4456$ 		& $50$ 	& $592$  	& $70.33 \pm 39.00$ s \\ \cline{2-7}
					       & C. Griseus 				& $4456$ 		& $6663$ 		& $30$ 	& $329$ 	& $89.41 \pm 3.47$ s \\ \cline{2-7}
					       & Mus Musculus 			& $8404$ 		& $13094$	& $50$ 	& $422$  	& $294 \pm 3$ s \\ \cline{2-7}
					       & Homo Sapiens 			& $8399$ 		& $13543$	& $100$ 	& $564$  	& $640^*$ s \\ \cline{2-7}
		 			       & Universal Model 			& $15638$	& $28301$	& $200$ 	& $2820$  & $2380^*$ s \\ \Xhline{2\arrayrulewidth}
		\multirow{7}{*}{$4$} & E. Coli 					& $72$ 		& $95$ 		& $20$ 	& $9$  	& $0.283 \pm 0.008$ s \\ \cline{2-7}
					       & Salmonella 				& $2436$ 		& $3357$ 		& $20$ 	& $53$  	& $17.28 \pm 0.16$ s \\ \cline{2-7}
					       & P. Tricornutum			 	& $2172$ 		& $4456$ 		& $50$	& $592$  	& $78.66 \pm 39.01$ s\\ \cline{2-7}
					       & C. Griseus 				& $4456$ 		& $6663$ 		& $30$ 	& $329$ 	& $97.46 \pm 3.47$ s \\ \cline{2-7}
					       & Mus Musculus 			& $8404$ 		& $13094$	& $50$ 	& $422$  	& $336 \pm 4$ s \\ \cline{2-7}
					       & Homo Sapiens 			& $8399$ 		& $13543$	& $100$ 	& $564$  	& $741^*$ s \\ \cline{2-7}
		 			       & Universal Model 			& $15638$	& $28301$	& $200$ 	& $2820$  & $3032^*$ s \\ \Xhline{2\arrayrulewidth}
		\multirow{7}{*}{$5$} & E. Coli 					& $72$ 		& $95$ 	 	& $20$ 	& $9$ 	& $0.234 \pm 0.023$ s \\ \cline{2-7}
					       & Salmonella 				& $2436$ 		& $3357$ 	 	& $20$ 	& $47$  	& $14.00 \pm 0.26$ s \\ \cline{2-7}
					       & P. Tricornutum			 	& $2172$ 		& $4456$ 	 	& $50$ 	& $544$  	& $60.95 \pm 21.36$ s \\ \cline{2-7}
					       & C. Griseus 				& $4456$ 		& $6663$ 	 	& $30$ 	& $306$ 	& $61.95 \pm 1.01$ s \\ \cline{2-7}
					       & Mus Musculus 			& $8404$ 		& $13094$ 	& $50$ 	& $383$  	& $378 \pm 121$ s \\ \cline{2-7}
					       & Homo Sapiens 			& $8399$ 		& $13543$ 	& $100$ 	& $528$  	& $584^*$ s \\ \cline{2-7}
		 			       & Universal Model 			& $15638$	& $28301$ 	& $200$ 	& $2653$  & $3124^*$ s \\ \Xhline{2\arrayrulewidth}
    	\end{tabular}
	\end{center}
	In this table, $m$ is the number of the metabolites (\emph{i.e.}, the number of rows in the $S$), $n$ is the number of the reactions (\emph{i.e.},  the number of columns in the $S$ or
	the number of rows in the $V$), and $c$ is the number of columns in the $V$, $L$ or $U$ (which is $1$ for the case that $v$ is a vector in first two rounds). The sparsity score,
	\emph{s.c.}, is equal to $\left \lVert v \right \rVert_1$ for round \hyperref[r:2]{2} and is equal to $\left \lVert V \right \rVert_{2,0}$ for rounds \hyperref[r:3]{3}, \hyperref[r:4]{4}, 
	and \hyperref[r:5]{5}. The gained score of each round in the competition has had a strong relationship with this parameter. Note that as round \hyperref[r:1]{1} is a feasibility problem, 
	sparsity means nothing there. \\
	In the datasets for round \hyperref[r:4]{4}, the parameters $\lambda$ are $7.125$, $251.775$, $133.68$, $333.15$, $392.82$, $203.145$, $212.2575$, respectively, 
	which force a huge penalty for releasing $Sv_j=0$ constraints and, therefore, have resulted in the same result with round \hyperref[r:3]{3} according to our heuristic \eqref{heu_d}, 
	which have made the set $J$ in \eqref{eq:J} to contain all the columns for all datasets. 
	In the datasets for round \hyperref[r:5]{5}, the $K$ parameters have been $4$, $4$, $10$, $6$, $10$, $20$, $40$, respectively. \\
	In our performance testing, the number of the iterations in the weighted algorithm has been set to $20$ for the round \hyperref[r:2]{2}, $10$ for the first six datasets in the rounds 
	\hyperref[r:3]{3}, \hyperref[r:4]{4}, and \hyperref[r:5]{5}, and $5$ for the BiGG universal model in the latter three rounds. Benchmarking has been done by the Julia library 
	\href{https://juliaci.github.io/BenchmarkTools.jl/dev/}{BenchmarkTools}, using $10000$ samples or any less number of samples during $300$ seconds limit of running (containing at least
	one sample). The aggregated results  in the table are included in the $\textit{mean} \pm \textit{std} \textit{ s/ms}$ format, or the $\textit{time}^* \textit{s}$ format for the huge datasets 
	on which only one sample has been taken. \\
	All codes have been run on a home MacBook Pro PC with a $2.2$ GHz Quad-Core Intel Core i7 processor, $16$ GB $1600$ MHz DDR3 memory, and Intel Iris 
	Pro $1536$ MB graphics. \\
	Additionally, a benchmarking for the preprocessing ideas, run on the round \hyperref[r:1]{1} datasets, is as follows:
	\begin{center}
    	\begin{tabular}{V{2.5} c | c | c V{2.5}}
   		\Xhline{2\arrayrulewidth}
    		Dataset & Running Time - No Process & Running Time - Processed \\ \Xhline{2\arrayrulewidth}
    		E. Coli 					& $0.37 \pm 0.65$ ms 	& $0.119 \pm 0.270$ ms \\ \hline
		Salmonella 				& $0.354 \pm 0.009$ s	& $7.563 \pm 3.607$ ms \\ \hline
		P. Tricornutum 				& $0.410 \pm 0.013$ s 	& $10.16 \pm 3.57$ ms \\ \hline
		C. Griseus 				& $1.273 \pm 0.012$ s 	& $25.44 \pm 9.85$ ms \\ \hline
		Mus Musculus 				& $4.758 \pm 0.160$ s  	& $72.02 \pm 33.97$ ms \\ \hline
		Homo Sapiens 				& $4.811 \pm 0.043$ s 	& $71.92 \pm 32.04$ ms \\ \hline
		Universal Model 			& $22.21 \pm 0.17$ s  	& $0.161 \pm 0.003$ s \\ \Xhline{2\arrayrulewidth}
    	\end{tabular}
	\end{center}
	which shows a significant improvement to the running time.
\section{Codes and date availability}
	Each round has contained three turns, which are three datasets to be judged. The Julia codes, written in Jupyter-notebook for each round (.ipynb file) and the data for all turns
	(folders T1, T2, and T3) for each round are provided in these URLs:
	\begin{itemize}
		\item Round $1$: \url{https://github.com/Optimizer-Competition-Pandas/Round_1}
		\item Round $2$: \url{https://github.com/Optimizer-Competition-Pandas/Round_2}
		\item Round $3$: \url{https://github.com/Optimizer-Competition-Pandas/Round_3}
		\item Round $4$: \url{https://github.com/Optimizer-Competition-Pandas/Round_4}
		\item Round $5$: \url{https://github.com/Optimizer-Competition-Pandas/Round_5}
	\end{itemize}
	Note that the \href{https://jump.dev/MathOptInterface.jl/v0.8.1/apimanual/}{MathOptInterface} package is used for parsing the optimization problems, and the \href{https://www.gnu.org/software/glpk/}{GLPK} is used as
	the linear programming solver.
\section{Acknowledgements}
	The authors gratefully acknowledge \href{http://soal.math.sharif.edu}{SOAL} lab for organizing this competition.
	Mojtaba Tefagh would like to thank Amir Daneshgar and Mohammad Hadi Foroughmand-Araabi for their insightful suggestions and constructive feedbacks.

\end{document}